\documentclass[12pt]{iopart}
\usepackage{graphicx}
\usepackage{xcolor,amsfonts}
\usepackage{setstack}
\usepackage{bm}
\usepackage{array,color}
\usepackage{iopams}

\def\bm{\boldsymbol}

\let\dsp=\displaystyle

\def\eg{{\it e.g.}}
\def\ie{{\it i.e.}}

\newcommand{\veps}{\varepsilon}

\begin{document}
\title{The Computation of Local Stress in ab initio Molecular Simulations}
\author{Xiantao Li}
\address{Department of Mathematics, The Pennsylvania State University, University Park, PA 16802, USA}
\ead{xxl12@psu.edu}
\vspace{10pt}
\begin{indented}
\item[]Oct 2018
\end{indented}
\begin{abstract}
 Motivated by the increasingly more important role of ab initio molecular dynamics models in material simulations,
 this work focuses on the definition of local stress, when the forces are determined from quantum-mechanical descriptions. Two types of
 ab initio models, including the Born-Oppenheimer and Ehrenfest dynamics, are considered. In addition, formulas are derived for 
both tight-binding and real-space methods for the approximations of the quantum-mechanical models. The formulas are examined via 
comparisons with full ab initio molecular simulations.
 \end{abstract}


\maketitle

\section{Introduction}
The last few decades have witnessed considerable interest in ab initio molecular dynamics (AIMD) as a simulation tool in chemistry and material science \cite{marx2009ab,marx1996ab,car1985unified}. 
Compared to classical molecular dynamics (MD), where the interatomic potential is constructed in advance, AIMD determines on-the-fly 
the forces on the ions from  quantum-mechanical (QM) degrees of freedom, which remain active throughout the computation. AIMD is particularly
suitable for problems where classical MD models are difficult to construct, \eg, alloys with multiple species and materials interfaces, and problems where there are significant 
changes in the chemical environment, \eg, bonding patterns. There have been great recent progress  in the development of efficient AIMD models, see the reviews  \cite{marx2009ab,tuckerman2002ab} for more details.   Various software packages have also been developed and widely used \cite{aradi2007dftb+,marques2003octopus,giannozzi2009quantum}.

  Due to the advent of 
growing computing powers, it is conceivable that AIMD will soon be applied to large-scale mechanical systems,  at which point, studying the elastic field induced by 
 lattice defects, boundary condition, or electric field,  would be of interest.  
 While computing the total stress in QM models is relatively straightforward, see the monograph \cite{martin2004electronic}, to our knowledge, the computation of {\it local stress} has not 
 investigated or implemented.

Meanwhile, for classical MD models,  there have been consistent mathematical framework for the definition and computational methods to define the local stress,  mostly motivated by the  Irvine-Kirkwood formalism \cite{irving1950}.    A popular approach is by Hardy \cite{hardy1982,hardy2002}, where the local density and momentum distributions are represented by certain smooth kernel functions. Following the Newton's equations of motion in the MD model,  fundamental conservation laws can be derived, and the local stress can be identified from the momentum balance. There has been tremendous recent progress toward the definition and computation of such local stress, see  \cite{tadmor2010,andia2006classical,Andia20056409,Chen2003359,Chen2003377,cormier2001,Costanzo2005533,costanzo2004notion,fu2013on,fu2013modification,lutsko1988,MuBe94,MuBe93,swenson1983,tsai1979,wu2015consistent,WuYaLi12,yang2012,Zhou2003,zhou2002,Zimmerman2010,zimmerman2004calculation} and the references therein for recent development.

 The primary goal of this paper is to establish the notion of local stress in AIMD models, where the electronic degrees of freedom are
 described by the density-functional theory \cite{Kohn1965} and its extensions \cite{gross1990time,ullrich2011time}. Due to the wide variety of AIMD models, we will focus on two popular methods, including the Born-Oppenheimer MD (BOMD) and the Ehrenfest dynamics. The BOMD approach involves solving the eigenvalue problem at each time step, while in Ehrenfest dynamics, the wave functions are determined by solving the time-dependent Schr\"{o}dinger equation (TDSE). In both cases, the
formulas for computing the total force on an ion are similar. Of particular importance for the definition of the local stress is the distribution of the force among each pair of ions.  We show that the force decomposition critically depends on how the quantum-mechanical models are discretized in space.   We discuss two specific implementations, the tight-binding (TB) approach, where the wave functions are approximation by linear combinations of atomic orbitals (LCAO), and the real-space approximations, \eg, finite-difference or finite element approximations.   We will show that the force decomposition is rather straightforward in the former approach, since the interatomic distance is naturally involved in the Hamiltonian matrix. However, for real space methods, since the grid points are not attached to ion positions, such force decomposition is rather non-trivial. 



The rest of the paper is organized as follows. In section 2, we review the AIMD models, as well as the definition of stress in classical MD models. The tight-binding approximation, along with
the stress definition, will be presented in section 3. In section 4, we discuss the real space approximation methods, and we present some numerical results.

\section{Ab initio Molecular Models and Fundamental Conservation Laws}

Throughout the paper, we will denote the ion positions by $R_I$, and the coordinate for the electronic wave functions by $\bm r.$ 
The elastic field will be defined in terms of the spatial variable $\bm x.$ We assume that the system under consideration 
consists of $N_{\mathrm{ion}}$ ions, and $N_{\mathrm{ele}}$ electrons.  

\subsection{Two AIMD models}

Let us start with the Newton's equations of motion for the ions,
\begin{equation}\label{eq: md}
\left\{
 \begin{array}{rl}
   \dot{R}_I & = P_I/m_I, \\
  \dot{P}_I &= -\frac{\partial}{\partial R_I} E_\mathrm{tot}, \quad I=1,2, \cdots, N_\mathrm{ion}.
  \end{array}\right.
\end{equation}
Here $m_I$ is the mass; $R_I$ and $P_I$ represent the coordinate and momentum of the nuclei, respectively.  
The main departure of AIMD from conventional MD models is that the potential energy $E_\mathrm{tot}$ are determined from an underlying quantum-mechanical description, 
instead of using a pre-determined interatomic potential. How the
electronic degrees of freedom are introduced, on the other hand,  leads to different formalisms of AIMD. We will consider two  models that have been widely implemented. Extensions to other models, \eg, \cite{car1985unified,niklasson2006time,alonso2011statistics,kuhne2007efficient,li2005ab,lin2013analysis}, especially
 the Car-Perrinello approach  \cite{car1985unified}, are likely to be straightforward. 
\medskip

\noindent{\bf Born-Oppenheimer MD.} 
First, we consider the Born-Oppenheimer MD (BOMD) model, where at each step, one  solves the 
eigenvalue problem, 
\begin{equation}\label{eq: eig}
  \hat{H} \psi_\ell = \veps_\ell \psi_\ell, \quad \ell=1, 2, \cdots, N_\mathrm{ele}.
\end{equation}
Again we denote the 
number of electrons by $N_\mathrm{ele}$.  

For the hamiltonian operator $\hat{H}$, we will consider the density-functional theory \cite{Kohn1965}, which uses a noninteracting system with an exchange-correlation function
that  reproduces the electronic density of the full, interacting system. The Hamiltonian operator $\hat{H}$ (in reduced units) consists of the kinetic energy, Hartree and exchange correlation, and the external potential,
\begin{equation}\label{eq: ham}
 \hat{H} = -\frac{\nabla^2}{2} + \hat{V}_H[n] + \hat{V}_\mathrm{xc}[n]+ \hat{V}_{ext}.
\end{equation}

We will denote the second and third terms, which are functionals of the electron density, as Kohn-Sham potential,
\begin{equation}\label{eq: vks}
  \hat{V}_{KS}=  \hat{V}_H[n] + \hat{V}_\mathrm{xc}[n].
\end{equation} 
The Hamiltonian operator $\hat{H}$ depends on the ions' positions, which contributes to the electronic structure as an external potential $ \hat{V}_{ext}$. This will be explained in more details in the next section. As a result, the eigenvalue problems need to be solved repeatedly, since the position of the ions is changing continuously. 
After each eigenvalue problem is solved, the total energy and force on each atom are computed as follows,
 \begin{equation}\label{eq: ef0}
 E_\mathrm{tot} = \sum_{\ell=1}^{N_\mathrm{ele}} n_\ell \veps_\ell,  \quad \bm f_I = - \frac{\partial E_\mathrm{tot}}{\partial R_I}, \quad I=1,2, \cdots, N_\mathrm{ion}.
\end{equation} 
 The coefficients $n_\ell$ are the occupation numbers.  
 
The coupled system (\ref{eq: eig}) and (\ref{eq: md}) are the basis of BOMD models, which can be interpreted as differential-algebraic equations (DAE).  The main assumptions is that the electronic states are instantaneously relaxed to ground states during the movement of the nuclei.  Typically,  the explicit formulas for computing the forces (\ref{eq: ef0}) are derived based on the Hellmann-Feymann theorem. We will discuss these formulas in the context of tight-binding and real-space approximations. 

\medskip

\noindent{\bf Ehrenfest Dynamics.} Another type of AIMD models is the  Ehrenfest dynamics, \eg, \cite{li2005ab}, where wave functions 
are determined from the time-dependent Schr\"odinger equations,
\begin{equation}\label{eq: schr}
  i\partial_t  \psi_\ell =  \hat{H} \psi_\ell, 
\end{equation}
in conjunction with the Newton's equations of motion for the ions (\ref{eq: md}). The coupled dynamics describe an ion-electron coupling that occurs continuously in time. For the time-dependent model, we consider the time-dependent density-functional theory  (TDDFT) \cite{gross1990time,runge1984density,ullrich2011time}.

In this case, the formula for computing the forces in (\ref{eq: ef0})  will be identically the same as in the 
static case, but for the reason that the wave functions and ion positions are independent variables .

\subsection{Definition of local stress based on momentum balance and force decomposition}

Since the Newton's equation (\ref{eq: md}) remains in all AIMD models, the fundamental conservation law of momentum still holds. More specifically, given the 
momenta and coordinates of the nuclei, we define the local momentum,
\begin{equation}
  \bm j(\bm x, t) = \sum_{I=1}^{N_\mathrm{ion}} P_I(t) \varphi \big(\bm x - R_I(t) \big).
\end{equation}

This is the starting point in Hardy's derivation of stress \cite{hardy1982}, which we will follow here. The function $\varphi$ is introduced as an approximation
to the dirac delta function in the original   Irvine-Kirkwood formalism \cite{irving1950}, and it is assumed to be non-negative with integral being 1.  
 More specifically, by taking time derivative, one gets,
\begin{equation}\label{eq: j-bal}
 \begin{array}{rcl}
  \partial_t \bm j &=& \dsp -\nabla_{\bm x} \cdot \Big[ \sum_{I=1}^{N_\mathrm{ion}} P_I(t)  \otimes  P_I(t) \varphi \big(\bm x - R_I(t) \big) \Big]\\
    & & + \dsp \sum_{I=1}^{N_\mathrm{ion}} \bm f_I(t) \varphi \big(\bm x - R_I(t) \big).\\
\end{array}
\end{equation}

The conservation of momentum at the continuum level asserts that
\begin{equation}\label{eq: cons}
  \partial_t \bm j  = \nabla_{\bm x} \cdot \bm \sigma ,
\end{equation}
with $\bm \sigma$ being the total stress tensor. So the goal is to express the right hand side of the equation (\ref{eq: j-bal}) into a divergence form. 

This will be true if the force $\bm f_I$ exhibits the following patterns
\begin{enumerate}
\item The force can be decomposed as: 
\begin{equation}\label{eq: f-fij}
  \bm f_I = \sum_{J\ne I} \bm f_{IJ}
\end{equation}
\item Each component is skew-symmetric, 
\begin{equation}\label{eq: asym}
\bm f_{IJ}= - \bm f_{JI}.
\end{equation}
\item $\bm f_{IJ}$ is local: It should decay quickly as a function of the interatomic distance.
\end{enumerate}

The first two conditions are critical (and sufficient) to ensure the conservation of momentum. It is worthwhile to point out, although it is trivial to many, that the fact that the force 
can be decomposed as in (\ref{eq: f-fij}) does not by any means imply that the interaction is pairwise. In fact, the force component $\bm f_{IJ}$ may depend on atoms other than
the two atoms $I$ and $J$. The last property is not mandatory. However, when the forces are highly nonlocal, the computation would be much more expensive, since many particles have to
be visited.

To see why such force decomposition is necessary, we can substitute (\ref{eq: f-fij}) into the second term on the right hand side of (\ref{eq: j-bal}),  which yields,
\[
\begin{array}{rcl} 
& &\dsp\sum_{I=1}^{N_\mathrm{ion}} \sum_{J=1}^{N_\mathrm{ion}} \bm f_{IJ} \varphi \big(\bm x - R_I(t) \big),\\
&=&\dsp\frac12   \sum_{I=1}^{N_\mathrm{ion}} \sum_{J=1}^{N_\mathrm{ion}} \Big[ \bm f_{IJ} \varphi \big(\bm x - R_I(t)  \big) +  \bm f_{JI} \varphi \big(\bm x - R_J(t)  \big) \Big] \\
&=&\dsp\frac12   \sum_{I=1}^{N_\mathrm{ion}} \sum_{J=1}^{N_\mathrm{ion}}  \bm f_{IJ} \Big[ \varphi \big(\bm x - R_I(t)  \big) - \varphi \big(\bm x - R_J(t)  \big)  \Big]\\
&=&\dsp -\nabla\cdot \frac12   \sum_{I=1}^{N_\mathrm{ion}} \sum_{J=1}^{N_\mathrm{ion}}  \bm f_{IJ} \otimes R_{IJ} B_{IJ}.
 \end{array}
 \]
Here in the last step, we used the fundamental theorem of calculus, and converted the difference between the two terms into a line integral along $R_{IJ}=R_I-R_J$,
\begin{equation}
  B_{IJ}(\bm x) = \int_0^1 \varphi \big(\bm x - R_J - \lambda R_{IJ}  \big) d\lambda.
\end{equation}
This derivation reveals a divergence term, from which one can identify the local stress,
\begin{equation}
 \bm \sigma =-\sum_{I=1}^{N_\mathrm{ion}} P_I(t)  \otimes  P_I(t) \varphi \big(\bm x - R_I(t) \big) - \frac12   \sum_{I=1}^{N_\mathrm{ion}} \sum_{J=1}^{N_\mathrm{ion}}  \bm f_{IJ} \otimes R_{IJ} B_{IJ}(\bm x).
\end{equation}
 It is clear that such definition is always up to a divergence-free term, which
makes it non-unique \cite{tadmor2010}. The momentum in the first term can be further decomposed into a mean ($\bm j(\bm x,t)$) and relative momentum  and they lead to the convection term, typically appears on the left hand side of the momentum balance equation, and a kinetic stress term, \eg, see the derivations in \cite{MuBe93}. But our focus will be on the potential part of the stress (the last term).

In the next two sections, we will discuss how the force decomposition can be obtained when the force is determined from
quantum-mechanical models, including the algebraic description (\ref{eq: eig}) and the dynamics model (\ref{eq: schr}).

\section{Force Decomposition and Local Stress in Tight-binding models}

\subsection{Tight-binding methods}

Tight-binding (TB) methods represent an important class of approximations that are constructed using basis functions centered at atoms, here denoted by $\phi_\alpha(\bm r-R)$. We also 
choose $\alpha \in occ(R_I)$ with $occ(R_I)$ indicating the set of local orbitals. The  wave function $\psi_\ell$ will be approximated by  a linear combination of the atomic orbitals (LCAO),
\begin{equation}
  \psi_\ell \approx \sum_I \sum_{\alpha \in occ(R_I)} c_{\ell,I\alpha} \phi_\alpha(\cdot,R_I) , \quad \forall \ell =1, 2, \cdots, N_e,
\end{equation}

The first step in implementing TB is assembling the overlap and Hamiltonian matrices, defined as follows,
\begin{equation}
\begin{array}{rl}
  S_{I\alpha, J\beta}=& \langle \phi_\alpha(\cdot,R_I) |\phi_\beta(\cdot,R_J) \rangle, \\
  H_{I\alpha, J\beta} = &\langle \phi_\alpha(\cdot,R_I) |\hat{H} | \phi_\beta(\cdot,R_J) \rangle.
\end{array}
\end{equation}
The matrix elements only depend on the relative position of the atoms; $R_{IJ}= R_I - R_J$.  Namely, we can write $S_{I\alpha, J\beta}:=M_{\alpha,\beta} (R_{IJ})$. 
It satisfies the symmetry condition,
\begin{equation}
  M_{\alpha,\beta} (R_{IJ})= M_{\beta,\alpha} (R_{JI}). 
\end{equation} 
In practice, they are precomputed in advance and represented in parametric forms. 

For BOMD, the next step is usually the eigenvalue problem. Using a projection into the subspace spanned by the local orbitals, one can reduce the eigenvalue problem into a finite-dimensional generalized eigenvalue problem,
\begin{equation}
  H \bm c_\ell  =  \veps_\ell S \bm c_\ell.
\end{equation}
Here $c_\ell$ is an eigenvector, with elements denoted by $c_{\ell,I\alpha}$

Let $C$ be the matrix that contains the eigenvectors as columns, then the diagonalization can be written simply as,
\begin{equation}\label{eq: eig0}
   H C = SC \Lambda. 
\end{equation}
The diagonal matrix $\Lambda$ contains the eigenvalues $\veps_\ell$'s. 

\subsection{The force decomposition}
In order to define the force on an ion, we first observe that the total energy, 
  $E_\mathrm{tot} = \sum_\ell \veps_\ell n_\ell,$
 can be written in terms of the Hamiltonian matrix and the eigenvectors,
\begin{equation}
  E_\mathrm{tot} = \sum_\ell \sum_{I\alpha} \sum_{j\beta} H_{I\alpha,J\beta} c_{\ell, I\alpha} c_{\ell,J\beta} n_\ell.
\end{equation}

By taking the derivative with respect to the ion position $R_I$, and using the orthogonality condition in (\ref{eq: eig0}), one can derive  the Hellmann-Feymann formula,
which  in this case,  is given by,
\begin{equation}
 \begin{array}{rl}
  \bm f_I = & - 2\dsp \sum_{J\ne I}  \sum_{\alpha \in occ(R_I) } \sum_{\beta \in occ{R_J}} \frac{\partial}{\partial R_I} H_{I\alpha,J\beta}  n_\ell c_{\ell,I\alpha}  c_{\ell,J\beta}^*  \\
    & - 2\dsp \sum_{J\ne I}  \sum_{\alpha \in occ(R_I) } \sum_{\beta \in occ{R_J}} \frac{\partial}{\partial R_I} S_{I\alpha,J\beta}  n_\ell \veps_\ell c_{\ell,I\alpha}  c_{\ell,J\beta}^*. 
\end{array}
\end{equation}
The derivative of the eigenvectors with respect to the ion positions does not appear due to the orthogonality conditions (\ref{eq: eig0}), but not because they are not dependent of them. 

This provides a natural decomposition,
\begin{equation}
\begin{array}{lr}
 \bm f_{IJ}=& \dsp - 2 \sum_{\alpha \in occ(R_I) } \sum_{\beta \in occ{R_J}} \frac{\partial}{\partial R_I} H_{I\alpha,J\beta}  n_\ell c_{\ell,I\alpha}  c_{\ell,J\beta}^*  \\
    & - 2\dsp \sum_{\alpha \in occ(R_I) } \sum_{\beta \in occ{R_J}} \frac{\partial}{\partial R_I} S_{I\alpha,J\beta}  n_\ell \veps_\ell c_{\ell,I\alpha}  c_{\ell,J\beta}^*. 
\end{array}
\end{equation}

This is exactly how the forces are evaluated in the DFTB+ model \cite{aradi2007dftb+}. The formulas can be simplified, by introducing the density-matrix $\rho$, and the energy density-matrix $\Gamma$,
\begin{equation}
 \begin{array}{rcl}
   \rho_{I,J}&= &\dsp \sum_{\alpha \in occ(R_I) } \sum_{\beta \in occ{R_J}}    n_\ell c_{\ell,I\alpha}  c_{\ell,J\beta}^* , \\
   \Gamma_{I,J} &=& \dsp \sum_{\alpha \in occ(R_I) } \sum_{\beta \in occ{R_J}}    n_\ell \veps_\ell c_{\ell,I\alpha}  c_{\ell,J\beta}^*.
 \end{array}
\end{equation} 
Then the force can be written as,
\begin{equation}
   \bm f_{IJ}= - 2\mathrm{tr} \big ( \rho_{I,J} \frac{\partial} {\partial {R_I}} H_{I,J} \big) -2 \mathrm{tr} \big( \Gamma_{I,J}\frac{\partial} {\partial {R_I}} M_{I,J}  \big). 
\end{equation}
The fact that $\bm f_{IJ}$ is skew-symmetric is also evident. 

\subsection{A numerical test}

To verify the desired property of the force decomposition, we consider a silicon nanowire with 512 atoms.  The dimension of this quasi one-dimensional system is 
$96.96\AA\times  10.87 \AA \times 10.87 \AA$. The nanowire is divided into 16 block along the longitudinal direction. By integrating the conservation law (\ref{eq: cons}) in a block, 
we have,
\begin{equation}\label{bal-1d}
  \partial_t \bm j_k + \bm \tau_{k+1/2} - \bm \tau_{k-1/2} =0. 
\end{equation}
Here $\bm j_k$ is the total momentum in the $k$th block. $\bm \tau$ is the projection of the stress in the longitudinal direction, \ie, $\bm \tau= \sigma \cdot (1,0,0)$, and it represents the traction between two adjacent blocks. 

\begin{figure}[htbp]
\begin{center}
\includegraphics[scale=0.15]{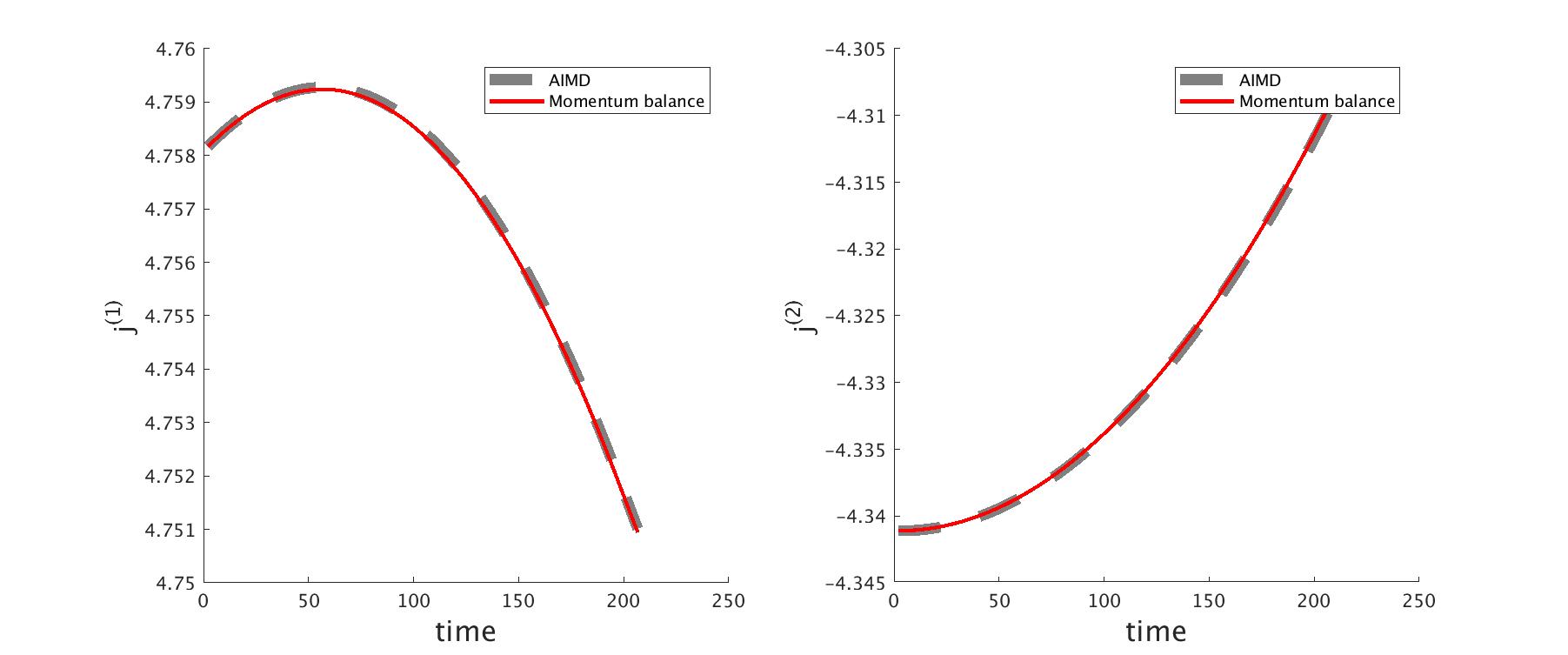}
\caption{Comparison of the total moment in a block of a nanowire, one computed from a BOMD simulation in DFTB+ (dashed curve), and the other computed 
from  the momentum balance  (\ref{bal-1d}) (solid line).}
\label{bal-tb}
\end{center}
\end{figure}

We follow the atomic units used in the software package, and modified the code to generate the force components $\bm f_{IJ}.$ 
We ran a BOMD simulation in DFTB+ for 100 steps with step size $\Delta t= 2.067$ (0.05 femto-second).  The initial velocity is randomly chosen.  Fig. \ref{bal-tb} shows the total momentum in a block in the longitudinal direction $j^{(1)}$, and a traverse direction $j^{(2)}(t)$. To examine the calculation of  $\bm f_{IJ}$, we computed the traction $\bm \tau$ from the AIMD simulation,  and then using the balance equation  (\ref{bal-1d}), we reconstructed the velocity at the same time steps.  Excellent agreement has been found. 

\section{Stress Calculation in Real Space Models}

In real space methods, the Hamiltonian operator will be discretized using finite difference  \cite{beck2000real} or finite element methods \cite{motamarri2013higher}. The Hamiltonian operator is then
represented at the grid points as a matrix, here denoted by $H.$ The numerical procedure in the quantum mechanical model would involve either 
solving an eigenvalue problem  (\ref{eq: eig}) or the time-dependent Schr\"{o}dinger equations (\ref{eq: schr}). The main difficulty is that the electronic degrees of freedom
are defined at grid point, and they are not specifically tied to the position of the ions.

\subsection{Force decomposition and stress calculations}
After the eigenvalues are computed, the total energy can be expressed as,
\begin{equation}
  E_\mathrm{tot}(R_1,R_2,\cdots,R_N)= \sum_{\ell} n_\ell \veps_\ell,
\end{equation}
with $n_\ell$'s being the occupation numbers.

The Hellmann-Feymann formula states that the change of the total energy with respect to the change of any parameter $\lambda$ is given by \cite{baroni_phonons_2001,gonze_dynamical_1997},
\begin{equation}
  \frac{\partial}{\partial \lambda} E_\mathrm{tot}  =   \sum_{\ell} n_\ell  \langle \phi_\ell | \frac{\partial \hat{H}}{\partial \lambda} | \phi_\ell \rangle.
\end{equation}

In real-space methods, the Hamiltonian operator is approximated at grid points. The only part of the Hamiltonian that explicitly depends on the nuclei position is the external potential,
\begin{equation}\label{eq: vext}
 \begin{array}{rl}
  V_{\mathrm{ext}}(\bm r)=&\dsp \sum_I \int w(\bm r - R_I) n(\bm r, t) d\bm r \\
  + &\dsp
  \sum_I\sum_{\ell} \int \int \psi_\ell(\bm r,  t)^* U(\bm r - R_I,\bm r' - R_I)  \psi_\ell(\bm r' ,t) d\bm r d\bm r'  \\
  + &    \dsp \frac12 \sum_{I=1}^N \sum_{J\ne I} \frac{Z_I Z_J}{|R_I - R_J|}.
  \end{array}
\end{equation}
Here $n(\bm r,t)$ is the electron density, given by,
\begin{equation}
  n(\bm r,t) = \sum_{\ell=1}^{N_e} n_\ell |\psi_\ell(\bm r, t)|^2.
\end{equation}

\noindent{\bf Remark:} It is worthwhile to point out that in principle, for ground state calculations, the electron density has an implicit dependence on the ion position. However, the Hellmann-Feymann theorem asserts that the derivatives only need to taken with respect to the $R_I$ in the exterma; potential. 

We will denote the three terms on the right hand side of (\ref{eq: vext}) as $V^\mathrm{loc}$, $V^\mathrm{nloc}$,  and $V^\mathrm{i-i}$, respectively.  The term $V^\mathrm{i-i}$ is a classical Coulomb interaction and the corresponding force decomposition is trivial, since it has a classical pairwise form and no explicit involvement of the wave functions:
\begin{equation}\label{eq: fii}
 \bm f^{\mathrm{i-i}}_{IJ}=  \frac{Z_I Z_J (R_I- R_J)}{|R_I - R_J|^3}.
\end{equation}

 Meanwhile, the first two terms stem from the pseudopotential approximation, \eg, norm-preserving potentials \cite{hamann1979norm}, which
 represents the influence of inner-shell electrons and reduce the problem to valence electrons. It represents the ion-electron interactions via a local term, here denoted by $w(\bm r)$, which usually only depends on the distance from an ion, and a nonlocal term, which consists of projections to local orbitals. Here we have represented the summation of the projectors by the function $U$. Interested readers are referred to \cite{hamann1979norm} for specific forms of these terms. The details of the force calculations in this case has been explained in the monograph \cite{kikuji2005first}.

It suffices to consider the derivatives of the first two terms with respect to the ion positions. The Hellmann-Feymann  formula suggests to define,
\begin{equation}\label{eq: fie1}
\begin{array}{lll}
\bm  f^\mathrm{loc}_I &=& \dsp   \int \nabla w(\bm r - R_I) n(\bm r) d\bm r, \\
\bm  f^\mathrm{nloc}_I &=& \dsp   \int \int \psi_\ell(\bm r,  t)^* \big( \frac{\partial}{ \partial  \bm r}  + \frac{\partial}{ \partial \bm r'})   U(\bm r - R_I,\bm r' - R_I)  \psi_\ell(\bm r' ,t) d\bm r d\bm r'. \\
\end{array}
\end{equation}

However, this formula does not reveal the interactions among the neighboring atoms, \ie, the decomposition $\bm f_{IJ}.$ It is tempting to separate the integrals by dividing the entire domain  into non-intersecting subdomains,  each of which contains a nucleus. However, our numerical tests suggest that the symmetry condition is not fulfilled. More importantly, in the nonlocal term, the projector $U$ only depends on points very close to the atom $R_I$. As a result, separating the integral would not yield a term that explicitly depends on a neighboring atom $R_J.$ In fact, when the second formula in (\ref{eq: fie1}) is implemented, the force becomes an on-site force, for which a force decomposition is not possible due to (\ref{eq: f-fij}).

This difficulty has prompted us to examine the formula that determines the force and derive alternative expressions. In principle, the wave functions depend on the ion position as well, due to the external potential. Such dependence is implicit in the Feymann-Heymann theorem, since it is cancelled when the equation is left-multiplied by an eigen-function. To gain more insight, we observe the explicit dependence of the external potential on the relative positions to the atoms, $\{\bm r-R_I, I=1, 2, \cdots, \},$  and we will write the wave function as follows,
\begin{equation}\label{eq: psil}
 \psi_\ell(\bm r,t) = \phi_\ell(\bm r-R_1, \bm r - R_2, \cdots, \bm r- R_{N_\mathrm{ion}},t).
\end{equation}
Similarly we write the electron density as,
\begin{equation}\label{eq: dn}
 n(\bm r, t) = d(\bm r-R_1, \bm r - R_2, \cdots, \bm r- R_{N_\mathrm{ion}},t).
\end{equation}

It is, however, not necessary to know the explicit forms of the functions $\phi_\ell$ and $d.$ They are introduced here only to show explicitly the dependence on the
ion positions. 
Now, we turn to the local force (\ref{eq: fie1}). Using integration by parts, we obtain,
\begin{equation}
\begin{array}{rl}
 \bm  f^\mathrm{loc}_I = & \dsp   \int \nabla w(\bm r - R_I) n(\bm r) d\bm r, \\
 =&\dsp - \int  w(\bm r - R_I)  \nabla n(\bm r, t) d\bm r \\
   =&\dsp  \sum_{J}  \int  w(\bm r - R_I)  \frac{\partial}{\partial R_J} n(\bm r, t) d\bm r.
\end{array}
\end{equation}

The last step is due to the form (\ref{eq: psil}) and (\ref{eq: dn}). This motivates us to define the force decomposition as follows,
\begin{equation}\label{eq: floc}
 \bm  f^\mathrm{loc}_{IJ} =\int  w(\bm r - R_I)  \frac{\partial}{\partial R_J} n(\bm r, t) d\bm r.
\end{equation}

Similarly, we can define the force decomposition for the nonlocal part,
\begin{equation}\label{eq: fnloc}
  \bm  f^\mathrm{nloc}_{IJ} = 2 \mathrm{Re}  \int \int \psi_\ell(\bm r,  t)^*  W(\bm r - R_I,\bm r' - R_I)   \frac{\partial}{\partial R_J} \psi_\ell(\bm r' ,t) d\bm r d\bm r'
\end{equation}

The remaining issue is how to determine the change of the wave functions. This is done by using the density-functional theory perturbation (DFTP) approach \cite{baroni_phonons_2001}. Let us denote the change of the wave functions, due to an instantaneous change of 
the position of an ion in one direction, by $\delta \psi$. For the eigenvalue problem that needs to be solved in BOMD, the perturbation yields,
\begin{equation}\label{eq: st}
 \begin{array}{l}
\dsp( \hat{H} - \veps_\ell  )\delta\!\psi_\ell + \int \frac{\delta \hat{V}_{KS}[n_0(\bm r)]}{\delta n(\bm r')} \delta n(\bm r') d\bm r' \psi_\ell +\delta \hat{V}_{ext} \psi_\ell  =0, \\
\dsp \int \psi_\ell^*(\bm r) \delta \psi_\ell (\bm r) + \psi_\ell(\bm r) \delta \psi_\ell^* (\bm r)  d\bm r =0,
\end{array}
\end{equation}
for  $\quad \ell=1,2,\cdots, N_\mathrm{ele}.$
Here $n_0(\bm r)$ is the ground state electron density, and $$\delta n(\bm r) =2 \mathrm{Re} \sum_\ell n_\ell \psi_\ell^* \delta \psi_\ell$$
will correspond to the term $ \frac{\partial}{\partial R_J} n(\bm r, t) $ in (\ref{eq: floc}).  $\hat{V}_{KS}$ is the
part of the Hamiltonian that depends on the electron density (\ref{eq: vks}).  The term \( \delta \hat{V}_{ext} \) refers to the
derivative of the external potential with respect to the ionic position. The second condition is to ensure the orthogonality of the wave functions. 

The equations form a linear system for the wave functions, known as the Sternheimer equation,  which can be solved via iterative methods. The coefficients of the linear system only depend on the ground states. So one does not need to solve the eigenvalue problem again.  Such an approach has been  an important route to determine phonon spectrum, polarizability, dielectric constants, etc \cite{yabana2006real}. 

At this point, it is clear that the formulas   (\ref{eq: floc}) and  (\ref{eq: fnloc}) would satisfy the property (\ref{eq: f-fij}). But the skew-symmetric property (\ref{eq: asym}) does not seem to be a direct consequence of our derivation. Here we offer a heuristic argument: suppose that one of the ions, $R_J$,  undergoes an infinitesimally small displacement. In the Sternheimer equation (\ref{eq: st}), $\hat{V}_{ext}$ would be a function of $\bm r - R_J.$  As a result, we assume that the perturbed  electron density  will also be centered around $R_J$. We assume that it is asymmetric with respect to  $\bm r - R_J,$ \eg, $ \delta n(R_J- \bm r ) = -  \delta n(
\bm r - R_J ) $.
  Therefore, we can write (\ref{eq: floc}) as  
\[
 \bm  f^\mathrm{loc}_{IJ} =\int  w(\bm r - R_I)   \delta n(\bm r - R_J) d\bm r. \]
The local potential $w$ only depends on the relative distance. So by changing variables, $\bm r - R_I = R_J - \bm r'$, we find that,
\[
 \bm  f^\mathrm{loc}_{IJ} =  \int  w( \bm r -R_I )   \delta n(R_J- \bm r ) d\bm r=- \bm  f^\mathrm{loc}_{JI}.\]
The same argument can be made toward the nonlocal part of the pseudopotential (\ref{eq: fnloc}) using a similar assumption on the density matrix.

\medskip
For Ehrenfest dynamics models, one can again take the derivative of the wave function with respect to the ion position. This procedure yields,
\begin{equation}
  i \partial_t \delta \psi_\ell = (\hat{T} + \hat{V}_{KS}[n_0] )  \delta \psi_\ell  +  \int \frac{\delta V_{KS}[n_0(\bm r)]}{\delta n(\bm r')} \delta n(\bm r') d\bm r' \psi_\ell +\delta \hat{V}_{ext} \psi_\ell.
\end{equation} 
In practical implementations, one can solve this linear Schr\"{o}dinger equation to determine $ \delta \psi_\ell,$ and implement the formulas (\ref{eq: fii}), (\ref{eq: floc}) and  (\ref{eq: fnloc}).

\subsection{A numerical test}

We consider a two-dimensional system -- a single layer graphene sheet with 32 atoms. The computation is done by using OCTOPUS \cite{marques2003octopus}, a real space implementation of the
ground state and time-dependent density-functional theory. Again, we use the standard atomic units. The length unit is Bohr radius and the energy unit is Hartree. All the results will be given
in terms of these two units. For the computation, the grid size in the finite-difference approximation is chosen to be $0.2066,$ and the simulation domain consists of  three-dimensional  $90 \times 78 \times 40$ grid with center of the rectangular domain shifted to the origin. The dimension of the computational domain is $18.595 \times 16.116 \times 8.265$ (in Bohr radius).   Periodic boundary conditions are applied in the first two space dimensions. We chose the standard pseudo-potential set in OCTOPUS. 

To create a non-homogeneous stress, we set the initial displacement as follows,
\begin{equation}
 \bm u(\bm x)= \frac12 \mathrm{e}^{-0.06 |\bm x|^2} \frac{\bm x}{|\bm x|}.  
\end{equation}
The atoms will be displaced according to this field in the first two dimensions with $\bm x$ as being the reference coordinates. We set the velocity to zero, so the stress is only due to the elastic field.  
We modified the part of the OCTOPUS code that computes the phonon spectrum. In particular, we followed the linear response calculations and computed $\delta \psi_\ell.$ The results are then used in
to compute the force decomposition in (\ref{eq: floc}) and  (\ref{eq: fnloc}).

\begin{figure}[htbp]
\begin{center}
\includegraphics[scale=0.15]{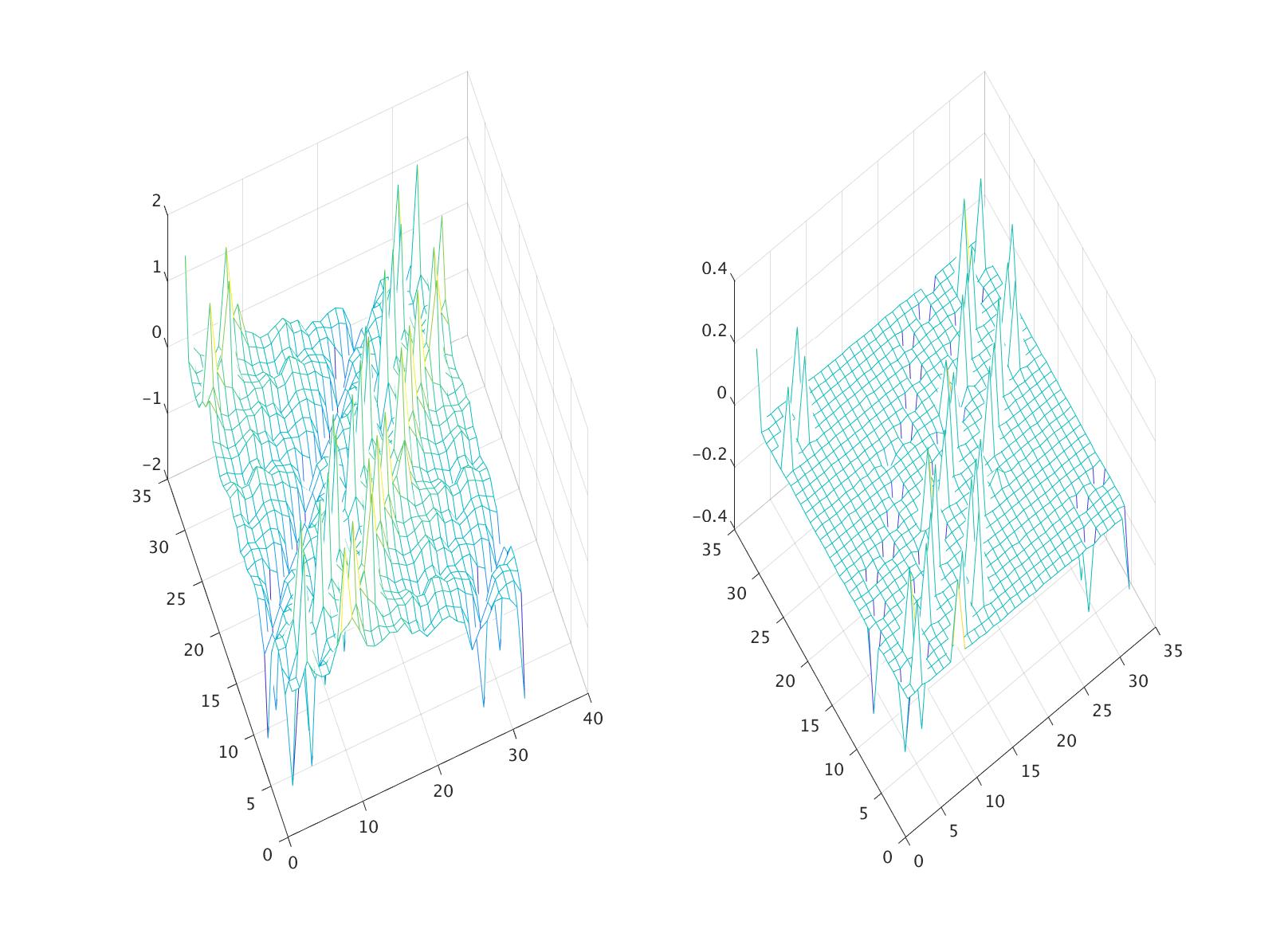}
\caption{The components of $\bm f_{IJ}$. Left panel: $\bm f_{IJ}^\mathrm{loc}$ from (\ref{eq: floc});   Right: $\bm f_{IJ}^\mathrm{nloc}$ from (\ref{eq: fnloc}). }
\label{fig: fij}
\end{center}
\end{figure}

In the first row of the table \ref{table: cmp}, we examine the force decomposition property (\ref{eq: f-fij}). More specifically, we add up $\bm f_{IJ}$ and compare the sum with $\bm f_I$, for each of the three contributions, (\ref{eq: fii}), (\ref{eq: floc}) and  (\ref{eq: fnloc}).  
In the second row, we verify the skew-symmetric property (\ref{eq: asym}), and again for each contribution.  More specifically, we compute the Frobenius norm of the matrix $\bm f_{IJ} + \bm f_{JI}. $ Even though the system has non-homogeneous deformation, one can see that the two properties (\ref{eq: f-fij}) and  (\ref{eq: asym}) are satisfied up to some small error. The error from the first row is clearly numerical error, since from the derivations of (\ref{eq: floc}) and  (\ref{eq: fnloc}), the condition (\ref{eq: f-fij}) should hold exactly. One may also choose to maintain the skew-symmetric property  (\ref{eq: asym}) exactly by defining $\tilde{\bm f}_{IJ} = (\bm f_{IJ} - \bm f_{JI})/2.$ In this case, the first condition (\ref{eq: fii})
 will hold approximately. Fig. \ref{fig: fij} displays plots of the matrices $\bm f^\mathrm{loc}_{IJ}$ and $\bm f^\mathrm{nloc}_{IJ}$ to provide a more direct view of the skew-symmetric structure.
\begin{table}[htp]
\begin{center}
\begin{tabular}{|c|c|c|c|} \hline
Force Contributions     & $\bm f^\mathrm{loc}$ &$ \bm f^\mathrm{nloc}$  & $\bm f^\mathrm{i-i}$   \\ \hline
$\dsp \frac{\sqrt{ \sum_I  |\sum_{J} \bm f_{IJ} - \bm f_I|^2 } }{ 
 \sqrt{ \sum_I \bm f_I^2} }   $    & 	0.0040	&  0.0043	& 1.2380$\times 10^{-9}$ \\ \hline
$\dsp \frac{\sqrt{\sum_{I,J} |\bm f_{IJ} + 
\bm f_{JI}|^2}  } {  \sqrt{\sum_{I,J} |\bm f_{IJ} |^2 }}$& 	0.0284	&  0.0734	& 5.1974$\times 10^{-23}$ \\ \hline
\end{tabular}
\caption{This table shows the results in the verification of the properties  (\ref{eq: f-fij})  and (\ref{eq: asym}).}
\label{table: cmp}
\end{center}
\end{table}

\begin{figure}[htbp]
\begin{center}
\includegraphics[scale=0.25]{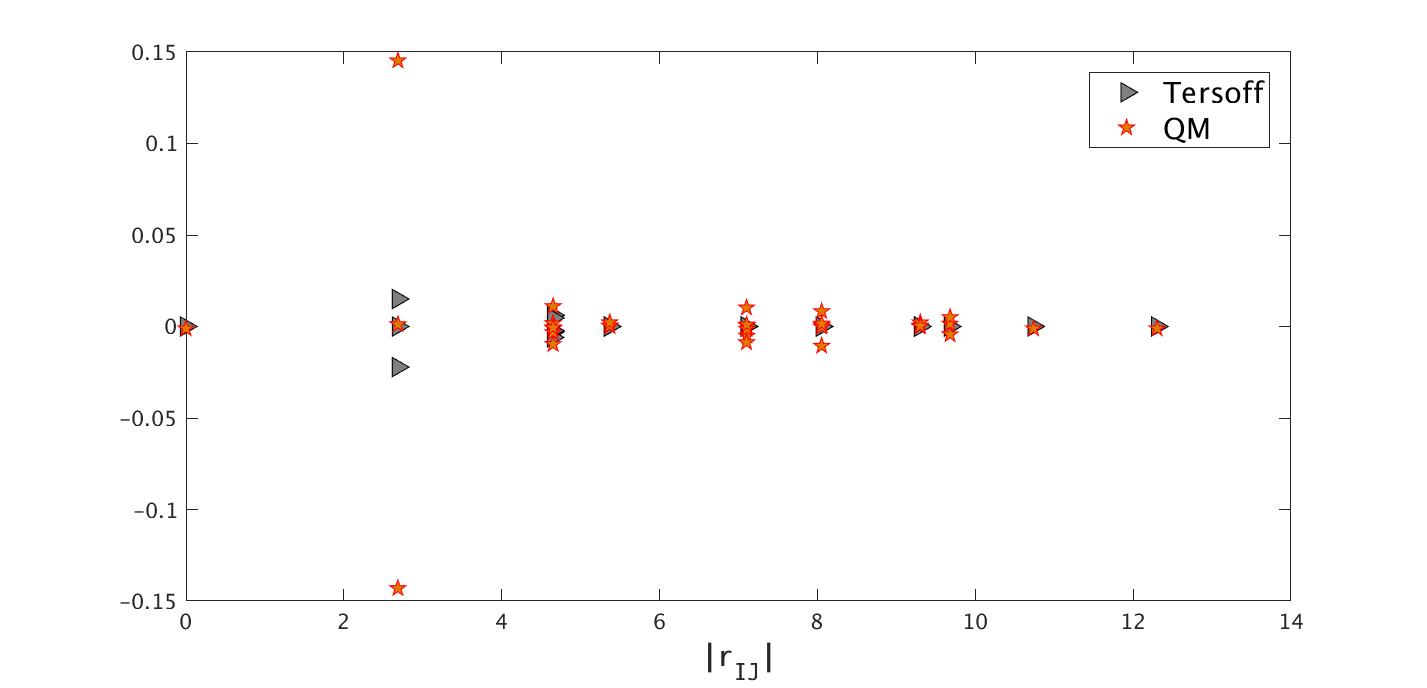}
\caption{The decay of the force components $\bm f_{IJ}$ and a function of the distance $r_{IJ}$ Also shown is $\bm f_{IJ}$ from the Tersoff potential.}
\label{fig: fijvsd}
\end{center}
\end{figure}

We now show in Fig. \ref{fig: fijvsd} how the magnitude of the force components $\bm f_{IJ}$  changes with respect to the distance between the two atoms ($r_{IJ}=|R_{IJ}|$). 
Interestingly, $\bm f_{IJ}$ exhibits a clear decay pattern, but it only becomes significantly smaller beyond the fifth neighbor, and negligible when the distance is around 10 Bohr (7th nearest neighbors). This is a lot larger than the cut-off radius of the Tersoff potential \cite{Te86}, also shown in the same figure. The force decomposition for the Tersoff potential  can be found in \cite{wu2015consistent}. In fact, there have been observations that the phonon spectrum of graphene will not be captured accurately, unless fifth neighbors included in the first-principle calculation \cite{mohr2007phonon}.    

Finally, we computed the local Hardy stress using the following two-dimensional kernel function, with a cut-off radius $ r_\mathrm{cut} =4$ Bohr,
\begin{equation}
 \phi(\bm x) = 
\frac{1}{\pi  r_\mathrm{cut}^2} \left\{\begin{array}{cc} 2 |\frac{\bm x}{ r_\mathrm{cut}}|^3 - 3 |\frac{\bm x}{ r_\mathrm{cut}}|^2 +1, & |\bm x| <r_\mathrm{cut} \\
 0,& \mathrm{otherwise.}  \end{array}\right.
\end{equation}

\begin{figure}[htbp]
\begin{center}
\includegraphics[scale=0.15]{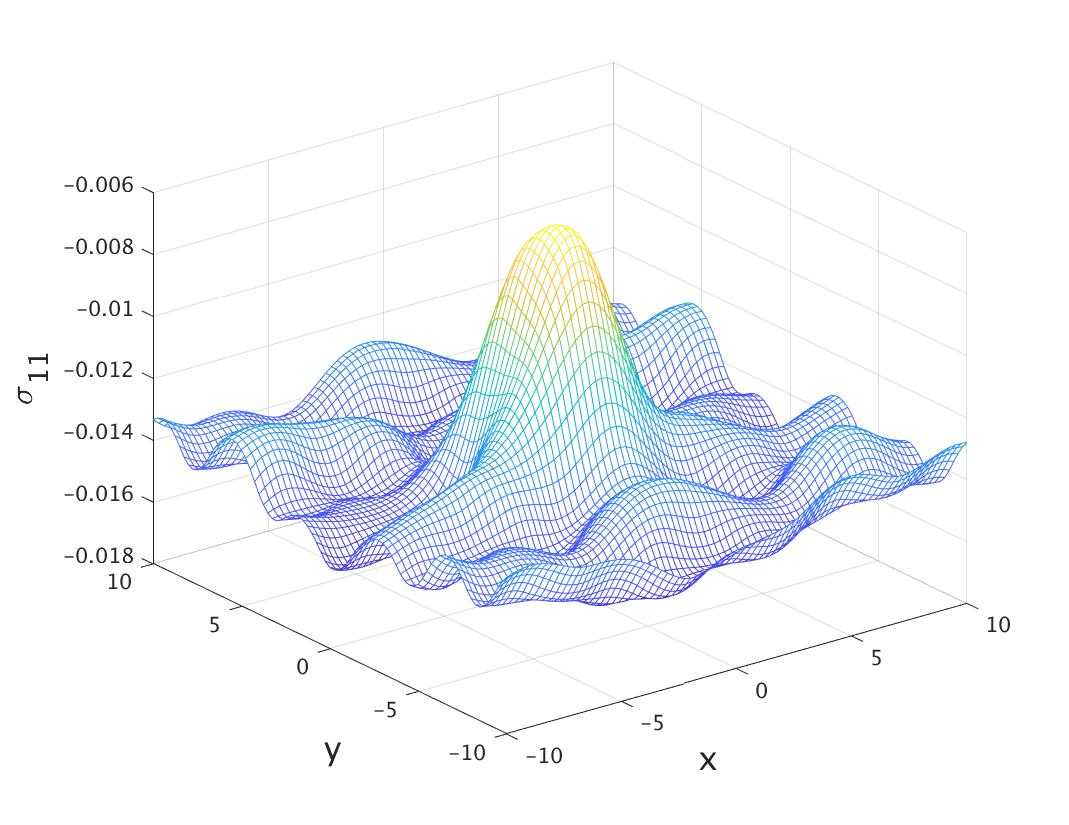}
\includegraphics[scale=0.19]{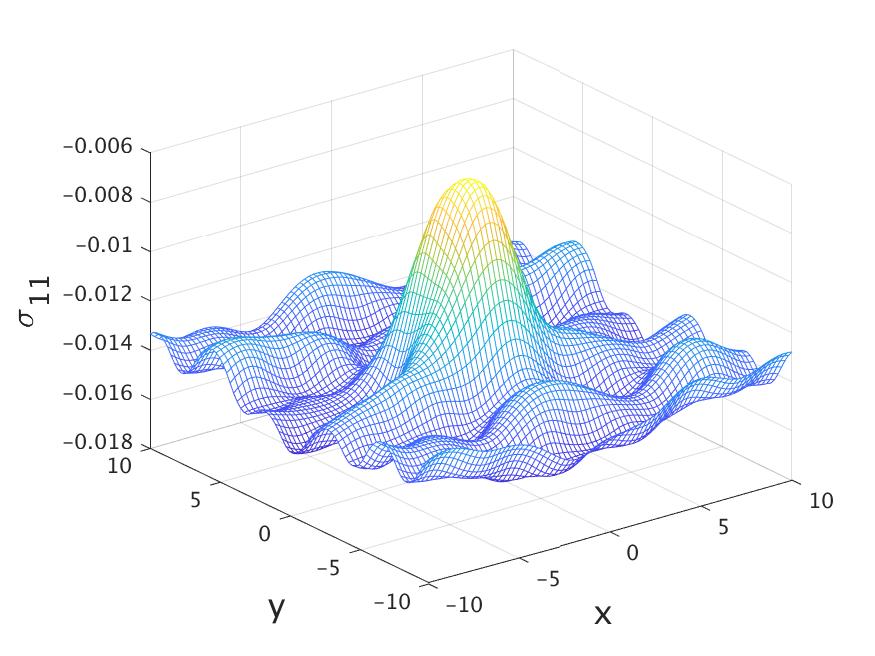}
\caption{The local stress $\sigma_{11}$ calculated from the Hardy's formalism using the force decomposition determined from
the DFT model (Left), and the classical MD model with the Tersoff potential. }
\label{fig: s11}
\end{center}
\end{figure}

We compared our results to that from the empirical Tersoff potential, and they exhibit very good agreement, as shown in Fig. \ref{fig: s11}. In principle, this should not be a means to validate the computational results. But in this particular case, since the displacement is quite smooth, we expect that the two results should be similar.  

\section{Summary and discussions}
We have investigated in this paper the appropriate force decomposition in ab initio molecular dynamics models. The goal is to be able to define the local stress that can either be used to analyze the elastic field, or be integrated to a continuum description to build a multiscale method. For tight-binding models, it turns out that such decomposition is rather straightforward. For real-space methods,  however, this becomes a non-trivial task.
We argued that one must estimate the change in the wave functions with respect to the ion displacement in order to obtain such a force decomposition. This perturbation can be computed within the linear response framework by solving the Sternheimer equation.

\section{Acknowledgments}
 This research was supported by NSF under grant DMS-1522617 and DMS-1819011. We would also like to thank the developers of DFTB+ \cite{aradi2007dftb+} and  OCTOPUS \cite{castro2006octopus} to make the software packages available to our group. 

\bigskip
\bibliographystyle{plain}

\end{document}